# Detection of confined current paths on oxide surfaces by local-conductivity atomic force microscopy with atomic resolution

C. Rodenbücher[1,2,*], G. Bihlmayer[2,3], W. Speier[2], J. Kubacki[4], M. Wojtyniak[4], M. Rogala[5], D. Wrana[6], F. Krok[6], K. Szot[1,4]

[1]Forschungszentrum Jülich GmbH, Peter Grünberg Institute (PGI-7), 52425 Jülich, Germany
[2]Forschungszentrum Jülich GmbH, JARA - Fundamentals of Future Information Technologies, 52425 Jülich, Germany
[3]Forschungszentrum Jülich GmbH, Institute of Advanced Simulation (IAS-1), 52425 Jülich, Germany
[4]University of Silesia, A. Chełkowski Institute of Physics, 40-007 Katowice, Poland
[5]University of Lodz, Faculty of Physics and Applied Informatics, 90-236 Lodz, Poland
[6]Jagiellonian University, Marian Smoluchowski Institute of Physics, 30-348 Krakow, Poland

*Correspondence to: c.rodenbuecher@fz-juelich.de

The analysis of the electronic surface properties of transition metal oxides being key materials for future nanoelectronics requires a direct characterization of the conductivity with highest spatial resolution. Using local conductivity atomic force microscopy (LC-AFM) we demonstrate the possibility of recording current maps with true atomic resolution. The application of this technique on surfaces of reduced $TiO_2$ and $SrTiO_3$ reveals that the distribution of surface conductivity has a significant localized nature. Assisted by density functional theory (DFT) we propose that the presence of oxygen vacancies in the surface layer of such materials can introduce short range disturbances of electronic structure with confinement of metallic states on the nanoscale.

Starting from the invention of the scanning tunneling microscope (STM) and the atomic force microscope (AFM) by Binnig et al. [1], scanning probe techniques have become indispensable tools in surface science in particular due to their ability of accessing various properties of the surface simultaneously such as topography, crystal structure or electrical conductivity [2, 3]. In recent years, the increasing demand for reducing the global consumption of electrical power from non-renewable resources has activated an intensive search for novel electrical materials turning out that transition metal oxides are promising candidates for the generation of electrical energy e.g. in solar cells or fuel cells as well as for building up computers with low power consumption consisting of redox-based random access memories (ReRAM) based on the resistive switching effect [4].

Since many of the exceptional phenomena of transition metal oxides are related to reactions taking place at the surface layer, the scanning probe technique is the method of choice in order to gain a deep insight into the fundamental physical processes on the nanoscale. In particular for the investigation of the resistive switching effect, the knowledge about the electronic transport phenomena in the surface layer is highly important. It is well known that the presence of extended defects such as dislocations leads to the evolution of electronic current paths with a diameter in the nanometer range as has been demonstrated using local-conductivity atomic force microscopy (LC-AFM) on the surface of the prototypical transition metal oxides $SrTiO_3$ and $TiO_2$ [5-9]. Recently, LC-AFM investigations of two-dimensional electron gases (2DEG) evolving at the

LaAlO$_3$/SrTiO$_3$ interface [10] attracted enormous attention showing that changes in local resistivity of several orders of magnitude confined to a few nanometers between a highly conductive and an almost insulating region could be successfully imaged by LC-AFM. In comparison to STM, the LC-AFM technique has the advantage that the measurement of the conductivity is independent from the measurement of the topography being realized by the detection of the atomic force in contact mode via the optical lever method. Hence it can be applied to surfaces with non-homogeneous conductivity using an ultra-small voltage in the mV range. However, while STM is well known to provide true atomic resolution on oxides for many years, the resolution of LC-AFM was limited to the nanometer range [11]. Only under certain conditions, periodic pattern with spacings in the range of the lattice constant were found also in current maps but they were related to a collective effect between a rather broad tip and the sample [12]. In contrast to this, in mapping the topography or friction force true atomic resolution has been achieved e.g. on ionic crystals [13-15]. It was also shown that with the help of Kelvin probe force microscopy (KPFM) the atomic-scale variation of the surface potential on TiO$_2$ (110) can be recorded [16]. In this paper we show that LC-AFM with true atomic resolution in the mapping of the current can be achieved. At first we demonstrate the principle of operation on the standard SPM calibration material - highly ordered pyrolytic graphite (HOPG). Next we analyze the local (sub-nanometer) distribution of surface electrical conductivity for the two prototypical transition metal oxides TiO$_2$ and SrTiO$_3$.

The experiments were conducted on different contact-mode AFM systems (JEOL and Omicron) equipped with Pt/Ir coated silicon tips (PPP-CONTPt, Nanosensors) and operated under UHV conditions at room temperature. A voltage was applied to the conducting tip and the current was recorded during scanning. The tips were mechanically and thermally stable allowing for the measurement of reproducible LC-AFM scans [17]. Although the total radius of the tip was in the range of 20 nm, the dimension of the contact point between the tip and the surface is in the sub-nanometer range [18, 19] (cf. detailed discussion in supplement). In order to demonstrate the performance of the tips on the atomic scale, we conducted measurements on freshly cleaved highly ordered pyrolytic graphite (HOPG). As shown in Fig. 1A, atomic resolution could be achieved in the maps of topography, friction force and current. In particular the current map reproduces trigonal arrangements of conducting spots as is well known from STM investigations of graphite [20]. The corresponding honeycomb structure is displayed as a guideline for the eye extracted from the topography map, in which the clearly visible hollow sites were taken as indicator for the arrangement of the lattice. Due to the crystal structure of graphite there are two non-equivalent atomic positions α and β resulting in a localization of the highest occupied electronic p$_z$ states close to the Fermi energy at the β site [21]. Hence it can be concluded that maxima of the conductivity in the LC-AFM map correspond to β sites while the minima of the friction force correspond to α sites. Since the AFM is operated in the contact mode, the friction force is strongly influenced by repulsive forces due to the Pauli Exclusion Principle being lower at the α sites. In order to verify that the observed LC-AFM map was obtained with "true" atomic resolution, produced a defective region on the surface by applying a voltage pulse (6 V) via the tip. After this we recorded the region again with low voltage (Fig. 1B). It can be seen that spots with different conductivity on the atomic range probably caused by surface point defects can be identified indicating that the conductance between tip and sample is confined to a sub-nanometer point contact proving that we are not dealing with collective effects such as the Pethica effect [22].

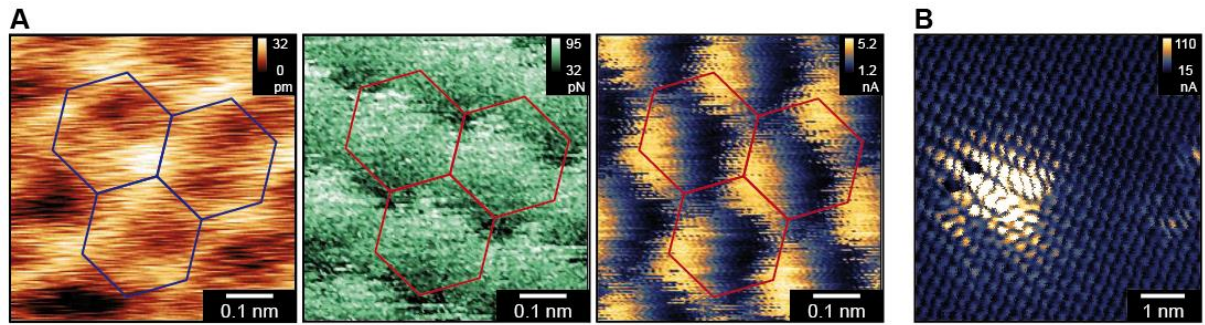

Fig. 1: A: Maps of topography, friction force and current measured by LC-AFM (U = 15 mV) on HOPG. B: Current map of an inhomogeneous region on HOPG demonstrating true atomic resolution.

Having seen that LC-AFM with atomic resolution can be obtained on homogeneous surfaces we now apply this method to the inhomogeneous transition metal oxides. We chose $TiO_2$, which can undergo an insulator-metal transition by the generation of oxygen vacancies making it a prototypical memristive material [6]. Since in the Ti-O system many different stable phases exist, complex transformations occur when $TiO_2$ is annealed under reducing conditions. Starting from a statistical distribution of oxygen vacancies, filamentary vacancy rows, extended Wadsley defects and finally suboxides such as Magnéli phases $Ti_nO_{2n-1}$ evolve with increasing non-stoichiometry giving rise to a inhomogeneous localized surface conductivity [23-25]. The $TiO_2$ (110) surface has been studied extensively by STM investigations which required cycles of sputtering and annealing to create well-ordered reconstructions and provide surface conductivity necessary for STM [26]. Since LC-AFM can be performed also on surfaces with non-uniform conductivity, we focused on the surface purely reduced at 800 °C for 1 h at a pressure below $10^{-8}$ mbar. Under these conditions, a macroscopic reconstruction or phase transformation did not occur as confirmed by low-energy electron diffraction (LEED) showing a perfect 1x1 symmetry (inset in Fig. 2A). On the other hand the reduction temperature was high enough to remove any carbon-based adsorbates from the surface (cf. supplement). Hence the reduced sample gave us the opportunity to catch up the first step of the surface transformation related to rearrangements of oxygen vacancies. In Fig. 2, the LC-AFM images of the reduced $TiO_2$ (110) surface obtained using a bias voltage of 10 mV are shown. The distribution of surface conductivity reveals a high degree of heterogeneity and not only conducting points marking the exits of single filaments but also agglomerations of conducting spots or clusters can be seen (Fig. 2A). In order to get an insight into the confinement of metallic states on the surface we zoomed in to an edge on a conducting cluster (Fig. 2B) where the highest gradient in conductivity was present. In the magnified current maps (Fig. 2C-E), the atomic arrangement can be clearly seen confirming that LC-AFM with atomic resolution is possible. Furthermore we can extract that the transition between a good conducting and an almost insulating surface region is spatially confined. The line profile of the local resistance calculated from the map in Fig. 2E displays that a drop of more than one order of magnitude in resistance takes place within only two or three lattice constants.

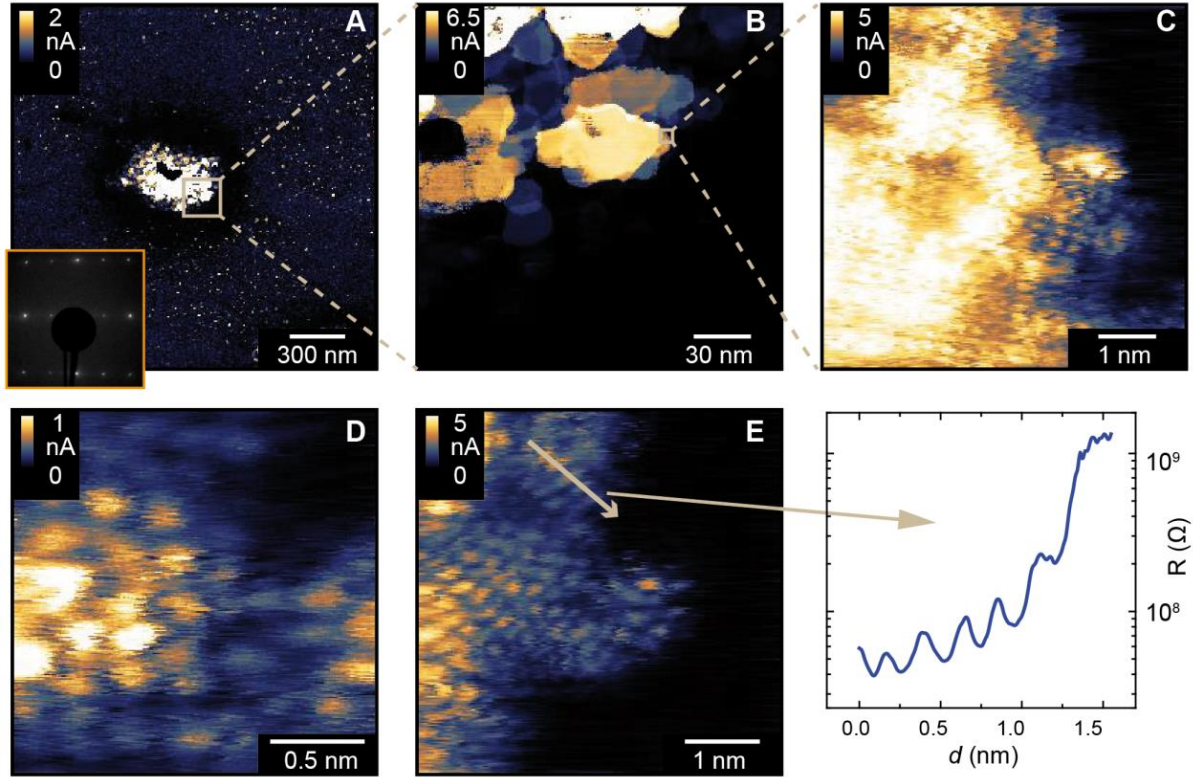

Fig. 2: Local conductivity measurements on the (110) surface of TiO$_2$ reduced at 800 °C (U = 10 mV). The inset in A shows the global LEED pattern.

Regarding the classical theory of doped semiconductors a confined clustering of the conductivity would not be expected. In order to gain a first insight how the measured local confinement of the conductivity could be generated, we performed calculations of the local density of states (LDOS) near oxygen vacancies in the TiO$_2$ surface region using the density functional theory (DFT). We simulated an oxygen vacancy at the surface and below the (110) surface in a 3×2 in-plane supercell of a film with nominal thickness of 10 unit cells in the ground state (for details see supplement). Apart from an in-gap state, the vacancy also induces some states at the bottom of the conduction band. Assuming that the current from the tip runs through these states which means that LC-AFM can be understood a mapping of the states close to the Fermi energy, we calculate the LDOS in a small energy window (50 meV) at the conduction band bottom. As shown in Fig. 3A, this density of states is highly localized in case of a single surface vacancy, which then could result in a locally confined conductivity as observed by LC-AFM. To illustrate this point further, we show the line-scan of the LDOS through the impurity in (1 -1 0) direction (Fig. 3C). In qualitative agreement with the measured line-scan in Fig. 2E we see an exponentially decaying density from the vacancy towards the stoichiometric surface. The density distribution depends sensitively on the Ti $d$-states present at the conduction band. The situation is even more involved if the oxygen vacancy is in the sub-subsurface layer at the same in-plane position (Fig. 3B): the complex hybridization pattern of the Ti $d$-states next to the vacancy produces near the surface a density distribution with maxima above the 5-fold coordinated Ti atoms appearing now between the surface oxygen rows. In this case, we observe already interaction effects between defect states of neighboring unit cells, nevertheless the extent of the charge-density perturbation decays on an nm-scale. Regarding not only single vacancies but extended defects such as a linear arrangement of vacancies it turns out that also in this case a local confinement of electronic states close to the defect is present [9]. Finally, in Fig. 3D, we simulated the LDOS of a vacancy cluster consisting of four surface and eight subsurface defects by superposition of Figs. 3A and 3B. In comparison with the experimentally observed LC-AFM image (Fig. 2D) we can see an agreement regarding distances between the conducting spots and confinement of electronic states which we confirmed by performing a statistic analysis of nearest neighbor distances between the local maxima revealing a value of $(0.29 \pm 0.05)$ nm for the experiment (Fig. 2D) and $(0.30 \pm 0.02)$ nm for the simulation. This shows that the assumption of clustered vacancies in the surface region could be a potential explanation for the observed localized conductivity.

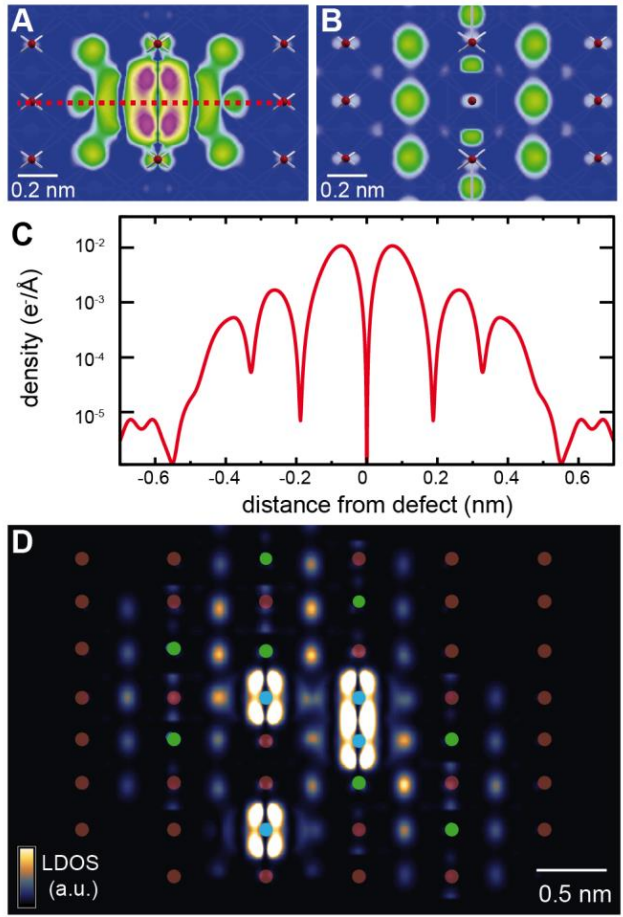

Fig. 3: Map of the local density of states at the bottom of the conduction band simulated by ab-initio calculations for an oxygen vacancy at (A) and below (B) the $TiO_2$ (110) surface. The topmost oxygen atoms are visible as red spheres; bonds to the subsurface Ti atoms are partially visible. The density is shown using a logarithmic scale with colors from blue ($10^{-4}$ e$^-$/Å$^3$) to red ($10^{-2}$ e$^-$/Å$^3$). The line profile (C) illustrates the spatial dimension of the density in configuration (A). (D) LDOS of a cluster consisting of four surface (blue) and eight subsurface (green) vacancies.

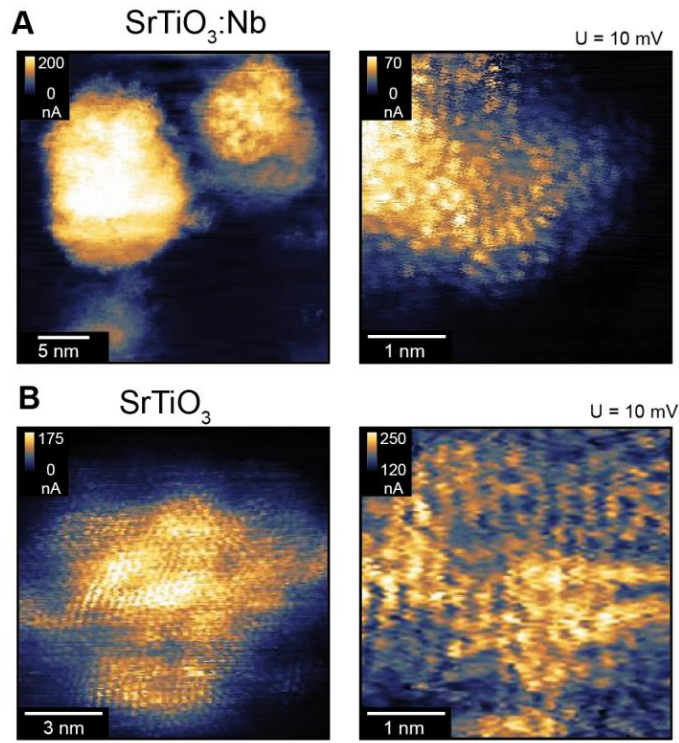

Fig. 4: LC-AFM measurements obtained with atomic resolution on the model perovskite $SrTiO_3$ doped with Nb (A), and without doping (B). Prior to the measurements the crystals were annealed in situ at 1000 °C.

Having seen that the surface conductivity of reduced $TiO_2$ has a distinct local nature, we extend the study to the second prototypical transition metal oxide $SrTiO_3$. We analyzed the (100) surface on donor (Nb) doped as well as on undoped crystals after reduction under UHV conditions [27]. In Fig. 4 it can be seen that on both surfaces atomic resolution in the current maps was achieved. On donor-doped $SrTiO_3$:Nb a cluster-like conductivity was found after reduction [8] which is expected from spectroscopic measurements [28]. Also on reduced undoped $SrTiO_3$ locally conducting areas were found and the periodic atomic arrangement could be resolved clearly in the current maps. It is striking that the local distribution of conductivity on reduced $SrTiO_3$ is relatively similar to that observed on reduced $TiO_2$ (cf. Fig. 2) which could be related to the fact that upon reduction the surface of $SrTiO_3$ becomes Ti-rich and hence could reflect the surface properties of $TiO_2$ [8, 29].

In summary we have shown that it is possible to record the current maps of materials with true atomic resolution using LC-AFM. We use this powerful technique to analyze the local conductivity of technologically prospective transition metal oxides $TiO_2$ and $SrTiO_3$ which is crucial for their applications. Heterogeneous distribution of conducting areas were found on the $TiO_2$ and $SrTiO_3$ revealing a local confinement of metallic states possibly related to a clustering of point defects such as oxygen vacancies. For certain defects it was illustrated with DFT calculation that single crystallographic defects can introduce a confinement of metallic states on the nanoscale which significantly affects the electrical transport in the materials. Our study shows that LC-AFM has a high potential to detect current paths in the wide class of oxide based electronic devices with true atomic precision.

We thank P. Meuffels and R. Waser for fruitful discussions. This work was supported in part by the Deutsche Forschungsgemeinschaft (SFB 917 "Nanoswitches").

**Supplement:**

I. The LC-AFM technique

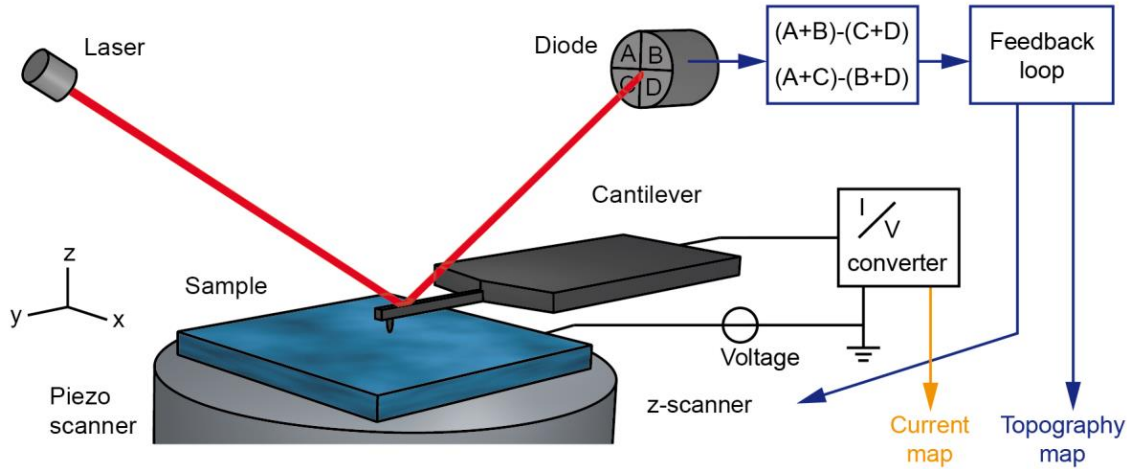

Fig. S1. Schematic of the LC-AFM technique

For the measurement of the local conductivity, a conventional AFM operated in contact mode is used. As illustrated in Fig. S1, the sample is positioned on a piezoelectric tube allowing for the movement of the sample in all three directions. A conducting tip on a cantilever serving as mobile top electrode is mounted in a fixed position and its deflection when being in contact with the sample surface is monitored by the optical lever method. Therefore a laser beam is focused on the cantilever and its reflection is directed to a four-sector diode. Via this optical amplification changes of the position of the tip can be detected with sub-nanometer resolution. Taking the measured intensity of the four sectors A, B, C, and D, either the out-of-plane movement of the tip related to the normal force $F_N$ or the in-plane-movement related to the lateral force $F_L$ can be extracted. The normal-force signal is fed into a feedback-loop keeping the force between tip and sample constant by regulating the extension of the piezo scanner in $z$ direction. In the contact mode used in the present study a repulsive force in the range of several nN was adjusted. Under these conditions it is well known that atomic resolution e.g. in friction force can be obtained without significant wear of tip or sample [2]. During acquisition of a LC-AFM scan the tip was rastered by moving the sample in $x$ and $y$ direction while the feedback loop was active. A mapping of the output signal of the loop moving the sample in $z$ direction results in the topography map. Additionally, a lateral force map related to the friction force can be acquired using the $F_L$ signal if the fast scan axis is perpendicular to the cantilever beam. In order to obtain information of the local conductivity, a voltage is applied between the sample contacted with metal clamps and the Pt/Ir coated tip. The current is measured using a sensitive current-voltage converter while scanning and mapped accordingly. Since ultra-small voltages in the range of several mV are used to measure the conductivity of the reduced oxide surface showing locally metallic conductivity, the influence of the offset voltage of the operational amplifier in the I/V converter has to be taken into account. Even in high precision amplifiers, the offset can be of several mV thus being in the same range than the externally applied voltage. Hence, a careful offset calibration has to be performed to minimize the error of the measurement. Processing of the obtained images was performed using the Gwyddion software involving background subtraction, contrast adjustment and drift correction. The quality and resolution of the LC-AFM map being an extremely surface-sensitive technique is determined by the electrical contact between tip and sample. To diminish the influence of contamination by adsorbates on the oxide sample we took great care to clean the surface by annealing under UHV conditions (see IV). Additionally, the annealing led to the generation of oxygen vacancies in the surface layer of the oxide crystals leading to enhanced conductivity due to electronic compensation. Due to the non-homogeneous surface conductivity found on the reduced oxides, a huge contrast in the local distribution of the current was present allowing to identify the conducting areas and the atomic resolution in particular at their borders easily.

II. Shape of the tip

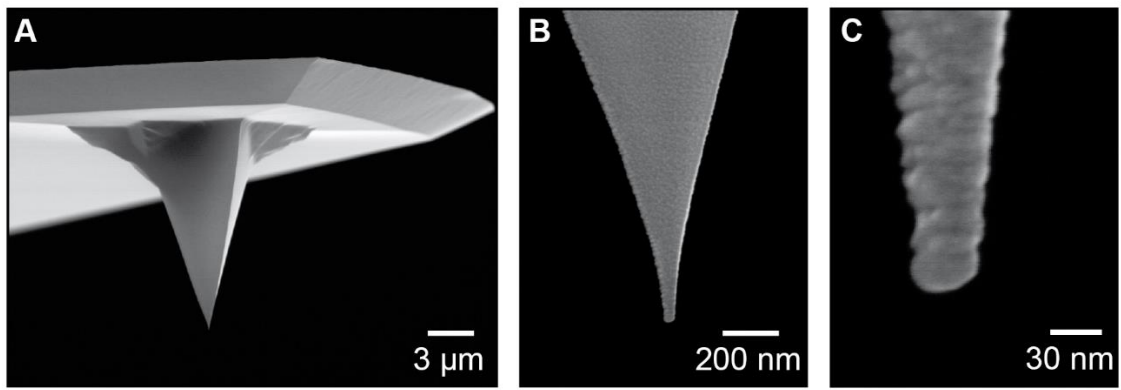

Fig. S2. SEM images of the Pt/Ir coated silicon tip used for LC-AFM

In order to gain insight into the nature of the contact between the Pt/Ir-coated silicon tip and the surface during current measurements with atomic resolution, we analyzed the tip shape by scanning electron microscopy (Fig. S2). The apex of the tip was relatively broad with a diameter in the range of 25 nm typical for a tip used for measurements in contact mode. The grain at the apex has a uniform shape and can thus be modelled as a regular sphere from which one atom can be expected to protrude enabling a monoatomic contact between the tip and the sample needed for LC-AFM measurements with atomic resolution as illustrated in Fig. S3E.

III. Investigation of HOPG

As standard material for scanning probe microscopy, HOPG was used and we were able to map single surface defects as described in the main text indicating the presence of true atomic resolution. Here, we intend to illustrate the variety in the measurement of local current that may help to understand the mechanism of local conductivity in more detail. As shown in Fig. S3A and S3B current images of two different types revealing either a hexagonal honeycomb structure or a trigonal structure can be obtained. This effect is known from STM investigations and it was concluded that the coupling between the topmost layer and the rest of the crystal is responsible for the different observed structures [30]. In case of a decoupled graphene sheet on top of the crystal a hexagonal symmetry is expected while the trigonal symmetry was found to be indicative for compact graphite [20]. Regarding the absolute current values in Fig. S3A and S3B it is striking that they are relatively low for a measurement of the direct current. Assuming that the current would be mainly determined by the resistance of the monoatomic point contact between the metallic tip and the HOPG having high in-plane conductivity, one would expect a resistance close to 13 kΩ which is the inverse of the conductance quantum [31]. This indicates that the measured current was not a direct current but a tunnel current which may be caused by an adsorbate layer being present on the tip as illustrated in green in the illustration in Fig. S3E. In this case the measuring mode would be similar to STM in constant height mode with optically controlled force feedback. To check this, we cleaned the tip by applying a voltage pulse of 5 V when the sample was in contact to the surface. After this we moved to another area and obtained an LC-AFM image shown in Fig. S3C. It can be seen that the current was significantly higher corresponding to an average resistance of 20 kΩ but atomic resolution with trigonal structure still could be observed. At the same time, the locally measured I/V curve (Fig. S3D) changed from a tunnel curve to a linear curve. This indicates that by cleaning the tip we removed the adsorbate layer from the apex of the tip and established a direct contact between the metal atoms and the atoms of the sample surface. Since the tip we used was coated with platinum, which is well known to be an active catalyst, the formation of the adsorbates cannot be avoided even under UHV conditions. On the other hand the existence of an adsorbate layer could also help to stabilize the tip when scanning in contact mode by enlarging the mechanical contact area and thus lower the local pressure (Fig. S3E).

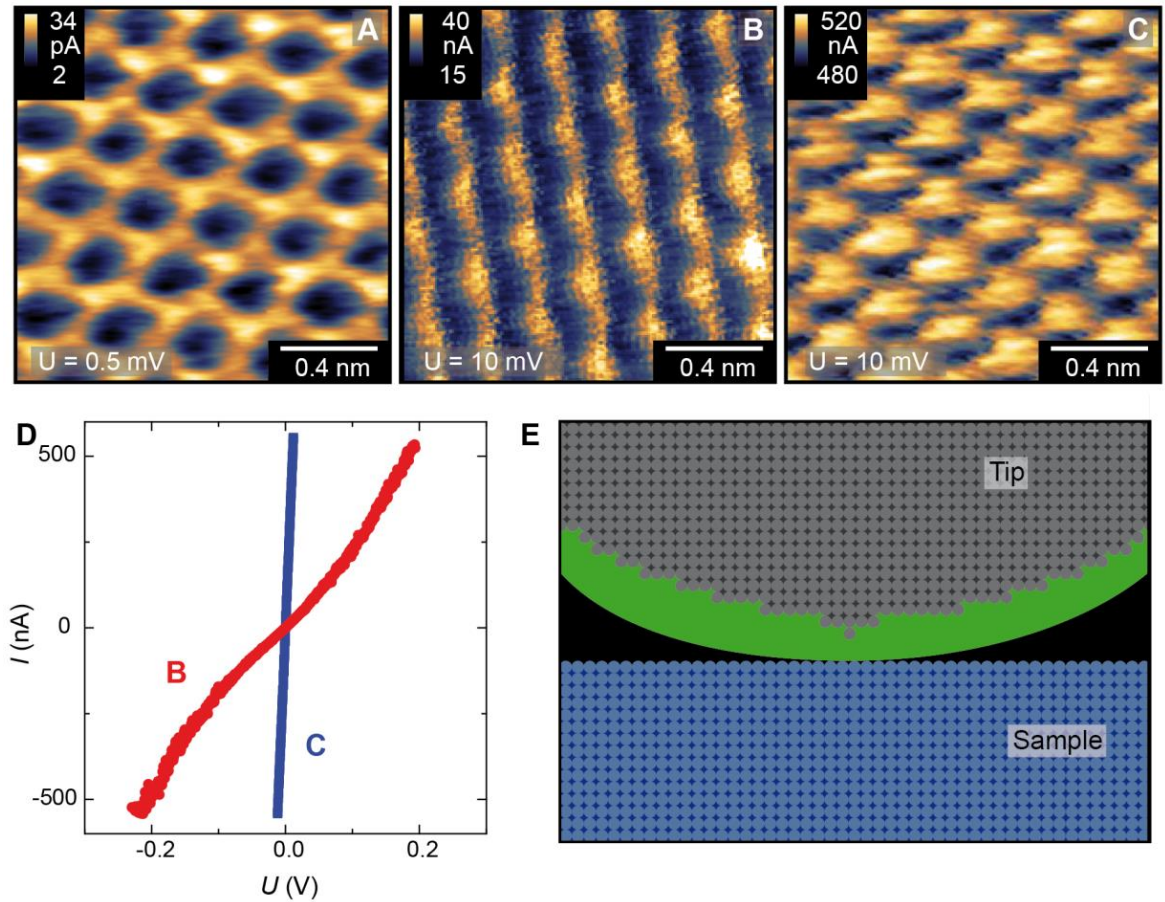

Fig. S3. LC-AFM maps obtained on HOPG using extremely low voltage of 0.5 mV (A) and low voltage (B). Image (C) was obtained using the very same tip as used in image (B) after tip cleaning by a voltage pulse of +5 V. (D) Local I-V curves corresponding to images (B) and (C). (E) Illustration of the contact between tip and sample in case of the presence of an adsorbate layer.

## IV. Chemical composition of the surface

The evolution of the chemical composition of the surface during annealing under reducing conditions was analyzed by X-ray photoelectron spectroscopy (XPS). The temperature was increased by steps of 100 °C from 500 °C to 900 °C with a dwell time of 1h in UHV before each measurement was started. In Fig. S4, the spectra of the core levels Ti2$p$, O1$s$ and C1$s$ measured on a TiO$_2$ (110) surface using excitation by X-rays from an Al K$\alpha$ source are presented. It can be seen that the TiO$_2$ spectrum consisted of a single doublet corresponding to the valence state 4+ and did not change significantly within the investigated temperature range. The O1$s$ spectrum was simulated by a main peak corresponding to lattice oxygen. Additionally a smaller peak at higher binding energies was present which decreased with reduction temperature and can be attributed to surface adsorbates. In the energy window of the C1$s$ line two peaks can be identified which were present only below a reduction temperature of 700 °C. Above this temperature, no carbon signal was measurable indicating that the cleaning of the surface from adsorbates by the thermal annealing was successful. From this measurement we can conclude that the inhomogeneous surface conductivity on the nanoscale as observed by LC-AFM (Fig. 2) is not related to carbon compounds but is a property of the reduced TiO$_2$ surface itself. Hence we assume that the presence of oxygen vacancies which could form clusters or precursors of linear defects such as Wadsley defects influence the spatial distribution of local surface conductivity strongly.

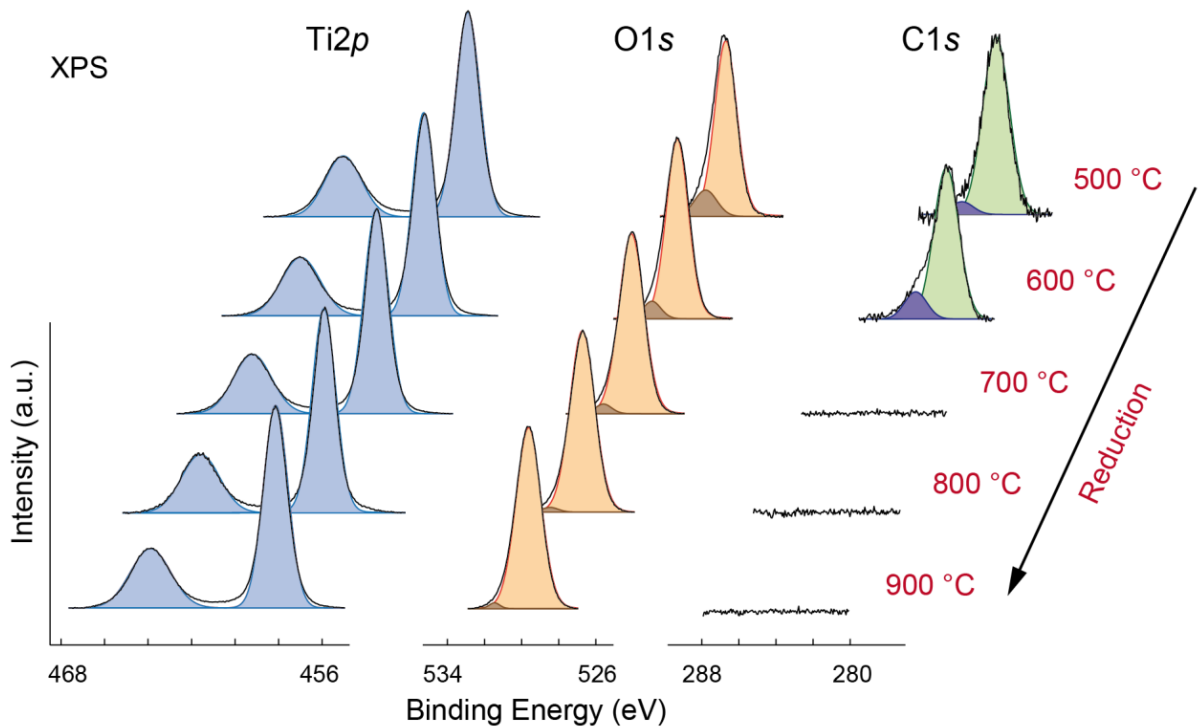

Fig. S4. Spectra of the core levels Ti2$p$, O1$s$, and C1$s$ of a TiO$_2$ (110) single crystal measured by XPS after annealing at different reduction temperatures under UHV conditions.

## V. Contact area of the tip

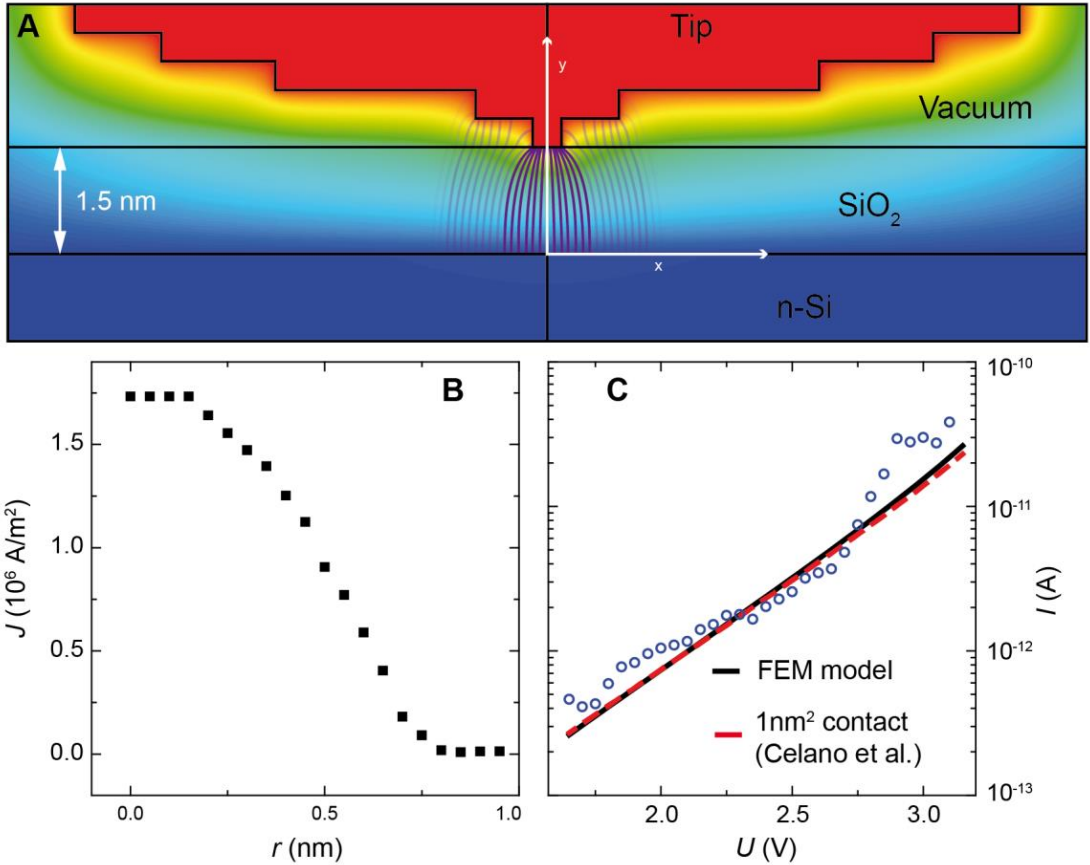

Fig. S5. FEM simulation of the contact between the AFM tip and a SiO$_2$ layer (A). Calculated current density as function of the distance from the center (B). Comparison of the measured tunnel current with our FEM model and the model presented by Celano et al.

As a convenient method for the estimation of the contact area between a tip operated in contact-mode and sample Celano et al. [19] have proposed the measurement of the tunnel current through a thin dielectric SiO$_2$ layer. According to the work of Frammelsberger et al.[18] this allows for the calculation of the effective contact area. Using this approach we characterized our tips on a 1.5 nm thick SiO$_2$ layer on top of n-doped Si. The measured tunnel current is shown as blue dots in Fig. S5C and corresponds to the expected tunnel current through an oxide with an effective electrode area of $A_{eff} = 1$ nm$^2$. Following the method of Celano et al. this theoretical current (red line in Fig S5C) was calculated using the formula Eq. 1 based on Fowler-Nordheim tunneling [18].

$$J = \frac{q^3 m_e}{16 \pi \hbar m_{ox}} \left(\frac{V_{ox}}{d_{ox}}\right)^2 \frac{1}{\Phi_B} \exp\left(-\frac{4}{3} \frac{\sqrt{m_{ox}}}{q \hbar} \Phi_B^{3/2} \frac{d_{ox}}{V_{ox}} \left(1 - \left(1 - \frac{q V_{ox}}{\Phi_B}\right)^{3/2}\right)\right) \quad \text{(Eq. 1)}$$

$$I = J \cdot A_{eff} \quad \text{(Eq. 2)}$$

Here, $J$ is the current density, $V_{ox}$ is the voltage across the oxide and $d_{ox}$ is the thickness of the oxide.
$m_{ox}$ is the electron mass in the SiO$_2$ layer and $m_e$ is the free electron mass. $\hbar$ is the reduced Planck constant, $q$ the electron charge, $\Phi_B$ is the barrier height and $I$ is the tunneling current.

In this description however, it is assumed that electron transport takes place exclusively in the circular contact area between tip and oxide. Since a LC-AFM measurement with atomic resolution would not be possible with such a large contact area of 1 nm², we argue in the following that also the assumption of a monoatomic tip-sample contact would be in agreement with the measured tunneling current. Therefore, we conducted a simulation using the finite element method (FEM) implemented in the Ansys software. As shown in Fig. S5A, we assumed a metallic tip with one atom at the apex following the geometric estimation of the atomic structure of a tip with 20 nm radius (cf. Fig. S1) being in contact with a $SiO_2$ layer. We calculated the potential distribution shown in false colors in Fig. S5A and the electron trajectories between tip and substrate shown in purple. For each trajectory we calculated the real distance that the electrons would have to travel during the tunneling event and calculated the current density as function of the distance $r$ from the center using Eq. 1 while the origin of our coordinate system was fixed to the interface between substrate and oxide. It can be seen in Fig. S5B that the maximum of the current density is located exactly below the tip-oxide contact but that there are also significant contributions at larger distances due to the curved shape of the electron trajectories. Additionally there is also a small contribution from electrons tunneling from the second atomic layer of the tip through the vacuum gap and the oxide to the substrate. For the sake of simplicity we only took into account the distance of these electron trajectories and still applied Eq. 1 and did not consider that these tunneling event would be in fact a two-layer problem. The contribution of tunneling from the second atomic layer of the tip to the total current is well below 10 % and thus does not play a significant role. To estimate the total current $I$ we calculated the current of each trajectory while due to the rotational symmetry of the problem, the corresponding areas for each trajectory have the form of a ring.

$$I = \sum_i J(d(r_i)) \cdot \pi(r_i^2 - r_{i-1}^2) \tag{Eq. 3}$$

The resulting total current is plotted in Fig. S5C as function of the distance from the origin in x-direction (black line). It can be seen that it is also identical to the calculation of Celano et al. for an effective emission area of 1 nm². This shows that our assumption of a monoatomic tip-sample contact can be brought into agreement to the measurement of tunnel currents through a dielectric. If we then would perform an LC-AFM measurement of a material with higher conductivity such as HOPG or the locally conducting oxides shown in the main text, the current would confined by the galvanic contact between the topmost atom of the tip and the sample allowing for atomic resolution. Additionally any tunneling current from the second layer of the tip would be orders of magnitude lower than the direct current and would not disturb the measurement. This is confirmed by analyzing local I-V curves measured when the tip was in contact to HOPG revealing an ohmic behavior with linear dependence between current and voltage after tip cleaning (cf. Fig. S3D).

In summary we have shown that a monoatomic tip-sample contact is not in contrast to estimation of the emission area during tunneling measurements which together with the fact that we could clearly resolve features on the atomic scale on HOPG lets us state that true atomic resolution in LC-AFM measurements is possible with standard tips used for contact-mode.

VI. DFT simulations

In order to gain insight into the spatial distribution of the local density of states close to oxygen vacancies in the TiO$_2$ (110) surface we performed calculations using DFT in the generalized gradient approximation [32] with corrections as proposed by Park et al. [33]. In this study, the full-potential linearized augmented plane wave method as implemented in the Fleur-code was used. We simulated oxygen vacancies at the surface and/or below the (110) surface in a 3×2 in-plane supercell of film with nominal thickness of 10 unit cells. The vacancies create in-gap states at the bottom of the conduction band. All atomic positions were allowed to relax resulting in ionic displacements that add to the screening of the electrons in these states. This leads, even at $T = 0$ K, to relatively well localized states characterizing the defects. The relaxation and electronic structure for the surface defect is very similar to the results obtained by Morgan and Watson [34]. As shown in the main text, we used the calculation of a single surface vacancy and a single subsurface vacancy to generate a larger vacancy cluster by superposition of the LDOS. As justification for this approach at first we calculated a cell containing both, a surface and a subsurface vacancy as shown in Fig. S6. The simulation shows in the upper panel the local density of states of a TiO$_2$ (110) surface with two oxygen vacancies in a region of 270 meV around the Fermi level. The corresponding 3×2 in-plane supercell is shown below with oxygen atoms indicated as large, red spheres and titanium as small, gray ones. Vacancy positions are marked with empty circles. The simulation shows that the LDOS can be regarded as linear superposition of the LDOS calculation of single surface and surface vacancies (see main text). This justifies the simulation of larger defect clusters by superposition, at least if the vacancies are not direct neighbors. This possibility was excluded in our thus generated image of the LDOS of the defect clusters.

The comparison of the LDOS, as obtained in DFT calculations, with the experimentally obtained LC-AFM conductance maps certainly needs some justification. In contrast to scanning tunneling microscopy, where this procedure is well known as Tersoff-Hamann model [35], in this method it is rather unrealistic to assume that the tip and sample wavefunctions are almost unperturbed in the course of the scanning procedure. Therefore, in the case where this perturbation is of the same order of magnitude as the conductance variation (e.g. on a metallic surface), such approximation will not be appropriate. In our case, i.e. a Pt tip in contact with a TiO$_2$ surface, the interaction can induce states in the conduction band of the sample [36] but this alone will not lead to local conductivity without the presence of oxygen vacancies that form pathways to the electrode [9]. Only when a connection to the vacancy-induced states in the surface is established, a significant current can flow. The resolution obtained in the LC-AFM map is, therefore, also limited by the length scale of the screening of the local field around the tip that is fortunately quite effective in the investigated oxides.

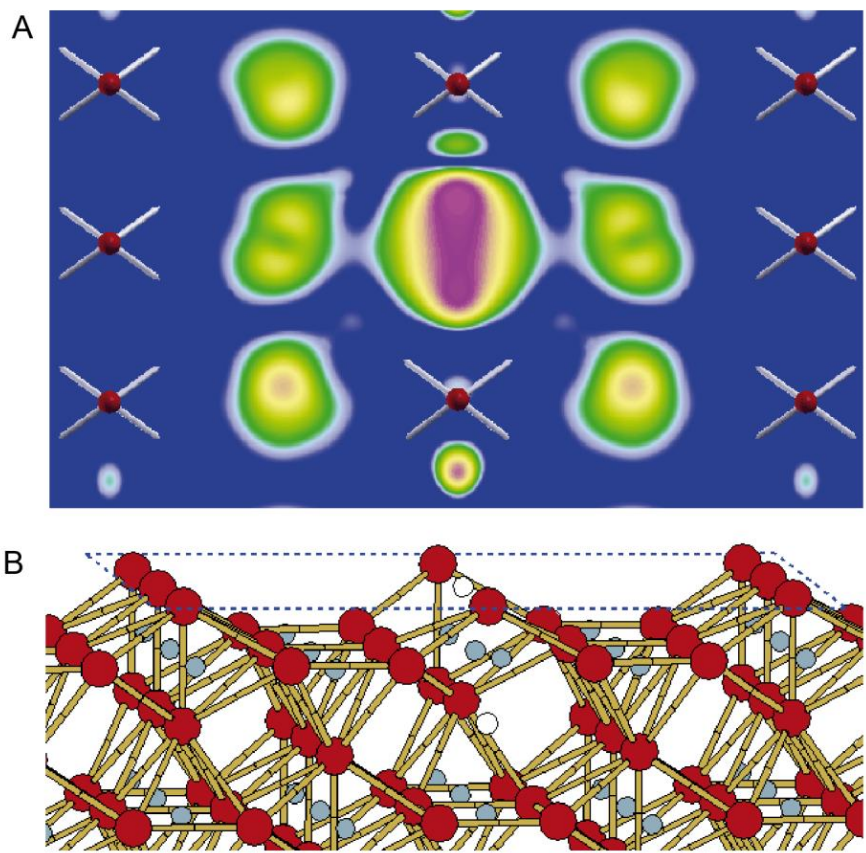

Fig. S6. (A) LDOS simulation of a TiO$_2$ (110) surface with one surface and one subsurface vacancy. (B) Corresponding ball-and stick model with Ti in gray and oxygen in red.